\documentclass[12pt,epsf,preprint]{aastex}

\shorttitle{The main Galactic radio-loops at 22 MHz}

\shortauthors{Uro{\v s}evi{\'c} and Borka Jovanovi{\' c}}

\begin{document}

\title{The brightness temperatures of the main Galactic radio-loops at 22 MHz}

\author{D. Uro{\v s}evi{\'c}\altaffilmark{1}, and V. Borka Jovanovi{\' c}\altaffilmark{2}}

\altaffiltext{1}{Department of Astronomy, Faculty of Mathematics,
University of Belgrade, Studentski trg 16, 11000 Belgrade, Serbia;
dejanu@math.rs, Tel: +381 11 20 27 827, Fax: +381 11 2 630 151}

\altaffiltext{2}{Atomic Physics Laboratory (040), Vin{\v c}a Institute of Nuclear Sciences, University of Belgrade, P.
O. Box 522, 11001 Belgrade, Serbia}

\keywords{radiation mechanisms: non-thermal –- radio continuum: general -- surveys}

\begin{abstract}
The average brightness temperatures and surface brightnesses at 22
MHz are derived for the four main Galactic radio-continuum loops
(Loops I, II, III and IV, hereafter radio-loops). Also the
radio-continuum spectra for the radio-loops are presented. Adding
the average brightness temperatures at 22 MHz derived here with the
average brightness temperatures derived from spectra published
previously at 408, 820 and 1420 MHz we obtained clearly non-thermal
spectral indices for all radio-loops. Our derived spectral indices
are slightly shallower than those measured by previous works.
\end{abstract}

\section{Introduction}

The radio-spurs are angularly immense and clearly visible features
on the radio-sky\footnote {The angular diameter of Loop I is
approximately $120^\circ$  [1].}. It is known that sets of
radio-spurs which form small circles on the celestial sphere are
called radio-loops. Four major Galactic radio-continuum loops were
recognized as Loops I - IV (see [2]). Their origin is probably from
local supernova explosions. They are probably very old (a few
million years) supernova remnants (bubbles) or superbubbles
originated in one or several supernova explosions, respectively. A
detailed review of the subject was published in [3].

In order to study the origin of emission from the radio loops, it is
necessary to determine their spectral indices. The spectral indices
of the radio-loops by using radio-continuum surveys at three
frequencies: 1420 MHz [4], 820 MHz [5] and 408 MHz [6], were studied
in [7], where then calculated the corresponding mean temperatures
and brightnesses. Until [7] and [8] papers there were no spectra
obtained by using mean temperatures for (at least) two different
frequencies, which is necessary for obtaining the simplest linear
spectrum. All earlier determinations of the radio loop spectral
indices were based on TT methods\footnote{The differential spectral
index for the extended radio object can be derived from measurements
at two different radio frequencies, when the brightness temperatures
are plotted at the so-called temperature -- temperature (T-T)
graph.}.

In this paper we derive the average brightness temperatures and
surface brightnesses of the radio-loops from the radio-continuum
survey at 22 MHz [9] and construct the new spectra of radio-loops
using data at the four different frequencies (1420, 820, 408, and 22
MHz).

\section{Analysis}

The data were obtained from the 22 MHz survey conducted with the
Dominion Radio Astrophysical Observatory (DRAO) [9]. The angular
resolution is 1.1$^\circ$ x 1.7$^\circ$ . The effective sensitivity
is about 500 K. The telescope comprises seven 9-m
equatorially-mounted paraboloids on an east-west line 600 m in
extent. These data are available on MPIfR's Survey Sampler
('Max-Planck-Institut f\"ur Radioastronomie', Bonn). This is an
online service (http://www.mpifrbonn.mpg.de/survey.html), which
allows users to pick a region of the sky and obtain images at a
number of wavelengths.

\begin{figure}[b!]
\centering
\includegraphics[width=0.7\textwidth]{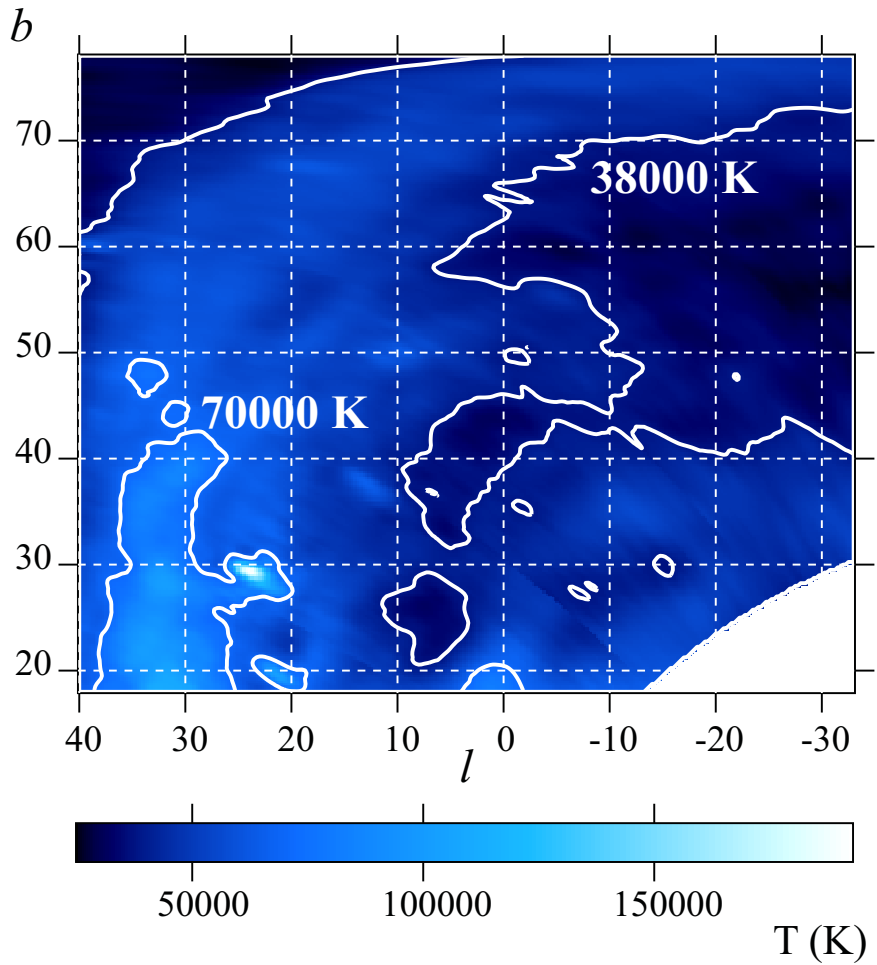}
\caption{Left: temperature scales for 22 MHz. It is given
in K and they are used for all figures of the radio-loops given below. Right: the area of Loop I at 22 MHz, showing contours
of brightness temperature. This contains the part of the North Polar Spur (NPS) normal to the
Galactic equator: l = [40., 0.]; b = [18., 78.], and its part parallel to the
Galactic equator: l = [360., 327.]; b = [67., 78.]. Two contours are plotted,
those representing the temperatures $T_{\rm min}$ and $T_{\rm max}$, as given in Table 1.}
\label{fig01}
\end{figure}

\begin{figure}[ht!]
\centering
\includegraphics[width=0.7\textwidth]{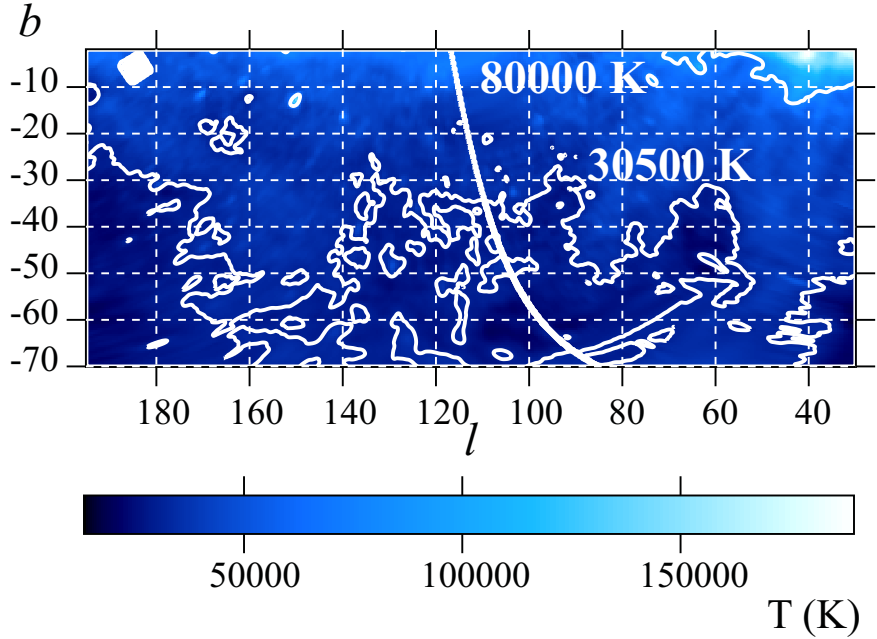}
\caption{The area of Loop II at 22 MHz, showing contours of brightness temperature. The two contours plotted represent the temperatures
$T_{\rm min}$ and $T_{\rm max}$, as given in Table 1. The white area and white curved line in the figure mean that no data exist there at that frequency. Spurs belonging to this radio-loop have
positions: l = [57., 30.], b = [-50., -10.] for spur in Aquarius and l = [195., 130.], b = [-70., -2.] for spur in Aries.}
\label{fig02}
\end{figure}

\begin{figure}[ht!]
\centering
\includegraphics[width=0.7\textwidth]{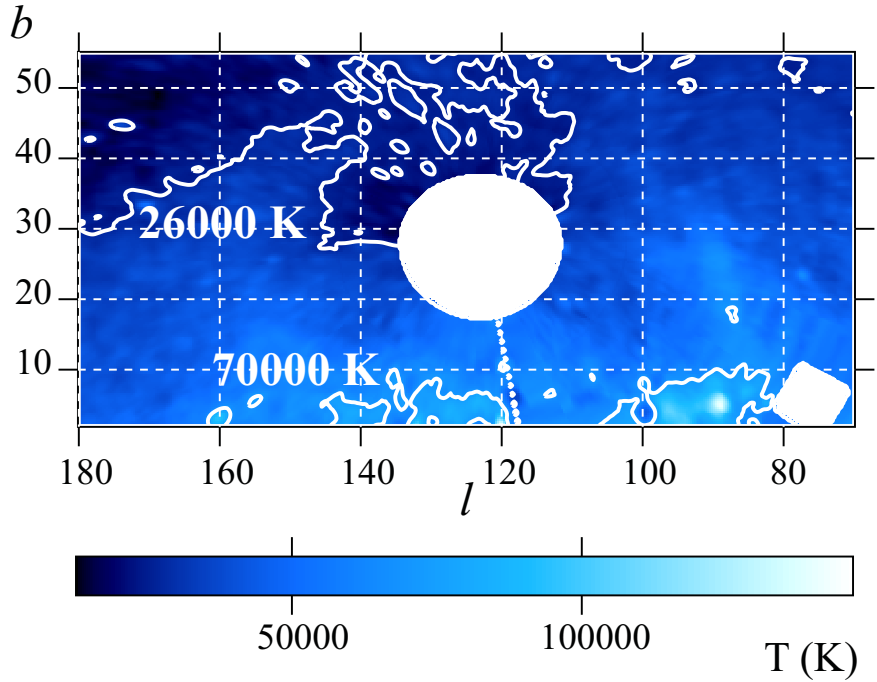}
\caption{The area of Loop III at 22 MHz, showing contours of brightness temperature. The two contours plotted represent the temperatures
$T_{\rm min}$ and $T_{\rm max}$, as given in Table 1. The white area in the figure means that no data exist there at that frequency. Spurs belonging to this radio-loop have positions: l = [180., 135.]; b = [2., 50.] and l = [135., 110.]; b = [40., 55.] for the first spur and l = [110., 70.]; b = [6., 50.] for the second one.}
\label{fig03}
\end{figure}

\begin{figure}[ht!]
\centering
\includegraphics[width=0.7\textwidth]{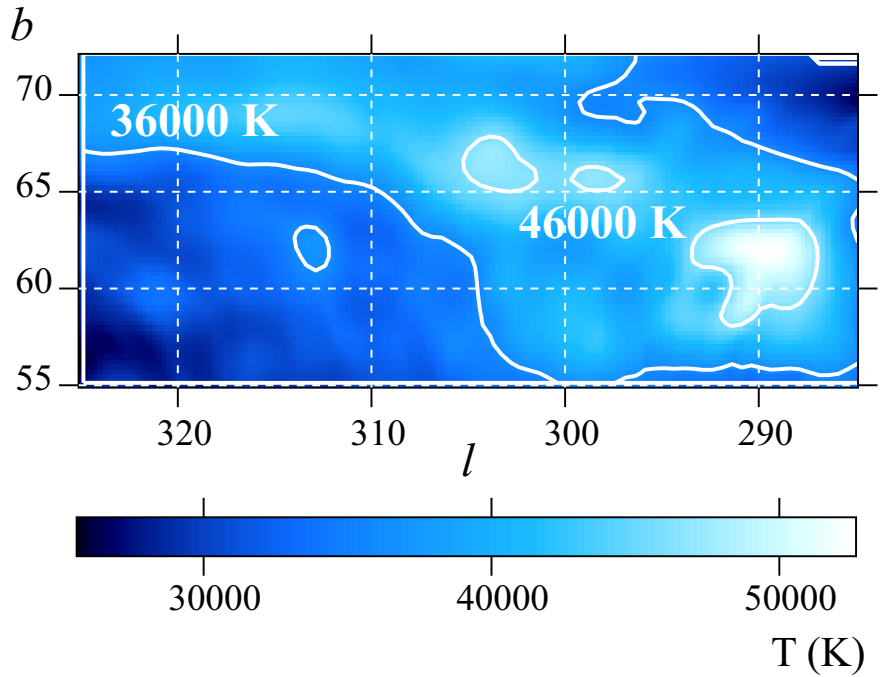}
\caption{The area of Loop IV at 22 MHz, showing contours of brightness temperature. The two contours plotted represent the temperatures
$T_{\rm min}$ and $T_{\rm max}$, as given in Table 1. This radio-spur has position: l = [330., 290.]; b = [48., 70.].}
\label{fig04}
\end{figure}

\begin{figure}
\centering
\includegraphics[width=0.7\textwidth]{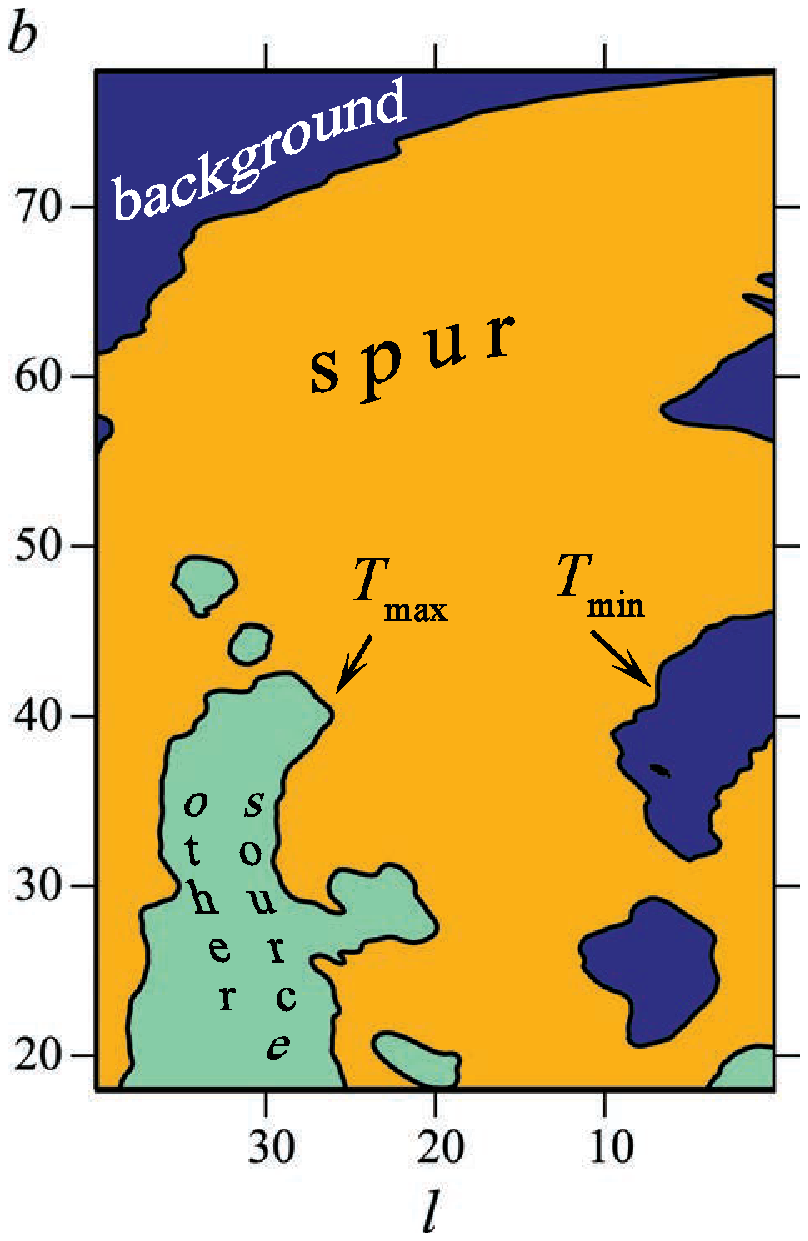}
\caption{The section of NPS from which data are sampled.}
\label{fig05}
\end{figure}

The areas of the loops were divided into different sections
(corresponding to spurs), and these sections were combined for the
final calculation of average brightness temperature a loop (see
Figs. (\ref{fig01} -- \ref{fig04})). The longitude and latitude
ranges which include spurs of the Loops I - IV are given in Table
\ref{tab01}. Background radiation was subtracted in the following
manner: first, the temperature of the loop plus background was
determined; next, the background alone immediately near the loop was
estimated (Fig. (\ref{fig05})); finally, we calculated the
difference of these values to yield the average temperature of the
loop.

\begin{table*}[h!]
\centering
\vspace{3cm}
\small\small
\caption{The Galactic longitude
and latitude for spurs belonging to Loops I -- IV and the lower and
upper temperature limits for these loops at 22 MHz}
\medskip
\begin{tabular}{l|lllll}
object               &  $l$ interval $(^\circ)$   &  $b$ interval $(^\circ)$     & $T_{\rm min}$ (K) &    $T_{\rm max}$ (K) \\\hline\hline
Loop I      & $l$ = [40, 0] & $b$ = [18, 78]  & 38000 &  70000  \\
            & $l$ = [360, 327] & $b$ = [67, 78]  &  &   \\
\hline
Loop II     & $l$ = [57, 30] & $b$ = [-50, -10]  & 30500&  80000  \\
            & $l$ = [195, 130] & $b$ = [-70, -2]  &  &    \\
\hline
Loop III    & $l$ = [180, 135] & $b$ = [2, 50]  & 26000&  70000  \\
            & $l$ = [135, 110] & $b$ = [40, 55]  &  &    \\
            & $l$ = [110, 70]  & $b$ = [6, 50]  &  &    \\
\hline
Loop IV     & $l$ = [325, 285] &  $b$ = [55, 72] & 36000&46000   \\
\hline
\end{tabular}
\label{tab01}
\end{table*}

The contour lines, which correspond to the minimum and maximum
brightness temperatures for each spur, are taken to define their
borders. $T_{\rm min}$ is the lower temperature limit between the
background and the spur, and $T_{\rm max}$ is the upper temperature
limit between the spur and unrelated confusing sources (superimposed
on the spur and hence requiring elimination from the calculation).
In this manner, background radiation was considered as radiation
that would exist if there were no spurs. We used averages over the
data within these two curves: the contour for $T_{\rm min}$  and the
contour for $T_{\rm max}$. More details are given in [7].
\newline
\newline
\newline

For evaluating brightness temperatures of the background, we used
all measured values below $T_{\rm min}$, inside the corresponding
intervals of Galactic longitude $(l)$ and Galactic latitude $(b)$,
and lying on the outer side of a spur (Fig. (\ref{fig05})). For
evaluating the brightness temperatures of a loop including the
background, we used all measured values between $T_{\rm min}$ and
$T_{\rm max}$ inside the corresponding regions of $l$ and $b$. Mean
brightness temperatures for spurs are found by subtracting the mean
values of background brightness temperature from the mean values of
the brightness temperature over the areas of the spurs. For more
details about the method see [7].

Generally, our estimates of the average brightness temperatures of
loops could be overestimated if the background temperatures have
some gradient.

\section{Results}

The results for the radio-loops obtained in this paper are presented
in Table \ref{tab02}, which lists the calculated average brightness
temperatures and surface brightnesses at 22 MHz, spectral indices
calculated using all, available average brightness temperatures at
1420, 820, 408, and 22 MHz. The last column in Table \ref{tab02}
shows the surface brightness $\Sigma$ at 1000 MHz calculated using
the spectral indices from third column in Table \ref{tab02}. The
surface brightnesses at 1000 MHz are calculated for comparison with
earlier results. When comparing results at 1000 MHz derived here
with values calculated earlier the good agreement of these results
is obvious (see [7] and references therein), and expected since the
measurement at 22 MHz is far from 1000 MHz to have any effect.

\begin{table*}[ht!]
\centering
\small\small
\caption{Brightness temperatures (K),
surface brightnesses $\Sigma$ $(10^{-22}{\rm ~W~m^{-2}~ Hz^{-1}~
sr^{-1}})$ and spectral indices $\beta$ ($T_{\rm
b}\propto\nu^{-\beta}$) derived in this paper. For comparison, in
the last two columns are given the spectral indices and the surface
brightnesses from Borka (2007) (signed by B07).}
\medskip
\begin{tabular}{l|rllllll}
object               &  \hfil $T_{\rm 22 MHz}$ \hfil   &  \hfil $\Sigma_{22 {\rm MHz}}$ \hfil & \hfil $\beta$ \hfil  & \hfil $\Sigma_{1000 {\rm MHz}}$ \hfil &\hfil $\beta$ (B07) \hfil  & $\Sigma_{1000 {\rm MHz}}$ (B07)  \\\hline\hline
Loop I      & 17600 $\pm$ 500 &$26.8\pm0.8$   & $2.64\pm0.03$&  $2.4\pm 0.6$&$2.74\pm0.08$  &\hfil $2.3\pm 0.7$\hfil\\
Loop II     & 12100 $\pm$ 500  &$18.3\pm0.8$   & $2.60\pm0.06$  &  $2.1\pm 0.7$&$2.88\pm0.03$ &\hfil $1.9\pm 0.5$\hfil \\
Loop III    & 18000 $\pm$ 500  &$27.3\pm0.8$   & $2.63\pm 0.02$ &  $2.4 \pm 0.6$&$2.68\pm0.06$ &\hfil$2.4\pm 0.6$\hfil\\
Loop IV     & \hfil 9400 $\pm$ 500   &$14.2\pm0.8$   & $2.77\pm 0.06$ & $0.8 \pm 0.6$&$2.90\pm0.30$ &\hfil$0.8\pm 0.7$\hfil\\
\hline
\end{tabular}
\label{tab02}
\end{table*}

\begin{figure}[ht!]
\centering
\includegraphics[height=0.4\textheight]{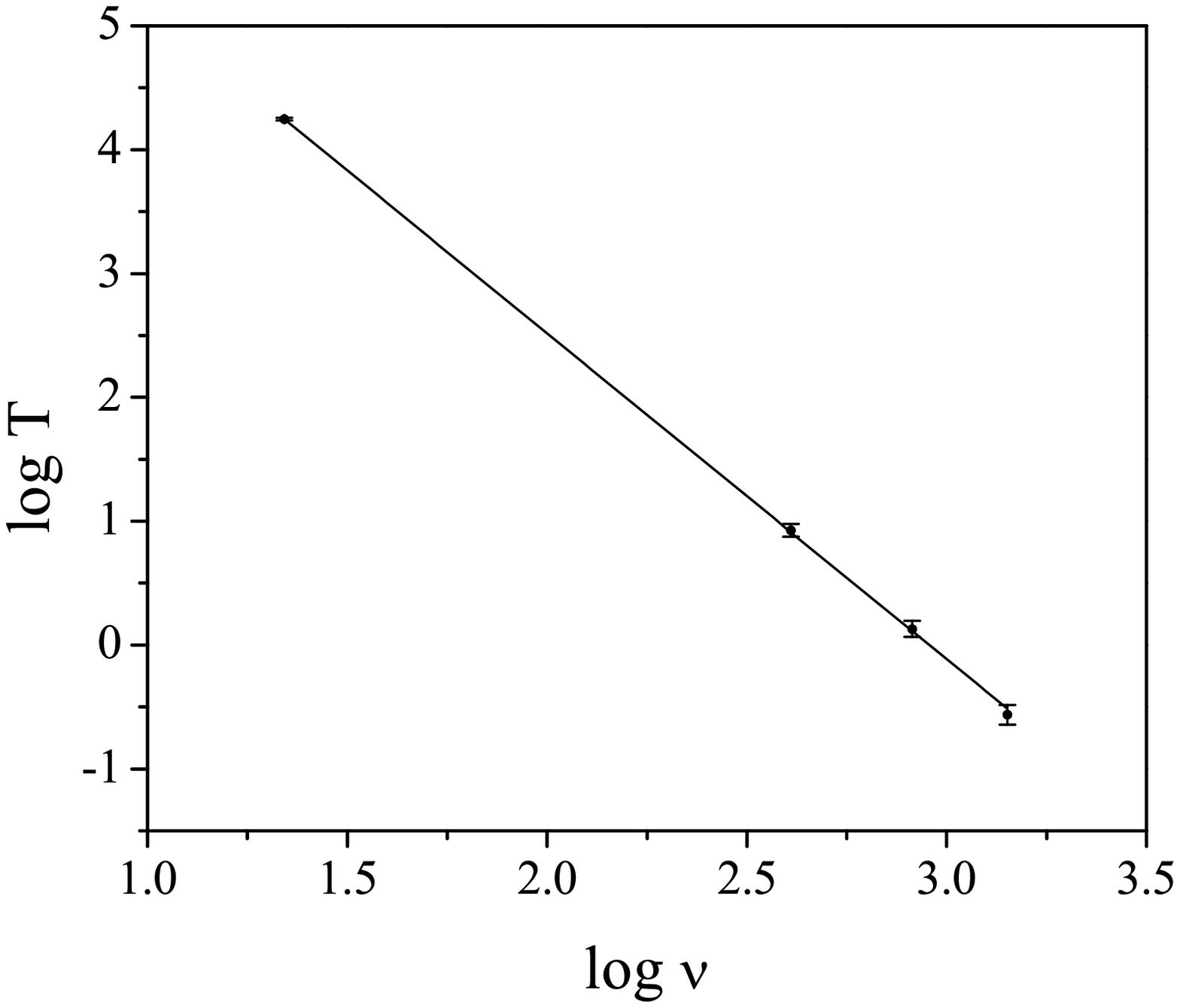}
\caption{The spectrum of Loop I: $\log T(\log\nu)$ for four measurements
– at 22, 408, 820 and 1420 MHz. Relative errors of the measurements
$\Delta\log T = {\Delta T\over T \ln 10}$ are presented by error bars, where $\Delta T$ are the corresponding
absolute errors given in Table 2.}
\label{fig06}
\end{figure}

Spectra of the radio-loops are presented in Figs.
(\ref{fig06}-–\ref{fig09}). If we compare the spectral indices
(defined by the relation $T\propto \nu^{-\beta}$) derived here with
spectral indices derived in [7] the slightly shallower slopes are
derived in this paper ($\Delta\beta\approx0.1$, see Table
\ref{tab02}). The radio-spectra of many astrophysical objects have
turnover near frequency of 100 MHz. The environment becomes
optically thick at low frequencies and due to this physical
condition, slope of the spectrum becomes opposite in its orientation
at the frequencies below turnover frequency. Anyway, we observed the
result of this effect as absorption in continuum (also it can be
synchrotron selfabsorption) in the low frequency part of
radio-spectrum. The measurements of surface brightness at 22 MHz
could be below the maximal surface brightness that should probably
be near 100 MHz. Since we added to the spectra the low frequency
points at 22 MHz, the shallower spectrum should be expected. The
shallower spectral indices obtained in this paper (Table
\ref{tab02}) should be used with caution -- the spectral indices
obtained earlier (with the data at higher frequencies) should be
preferable used for description of the synchrotron emission
originated from radio-loops. For eventual identification of spectral
turnover, the average surface brightnesses of the radio-loops
between 22 and 408 MHz are needed.

\begin{figure}[ht!]
\centering
\includegraphics[height=0.4\textheight]{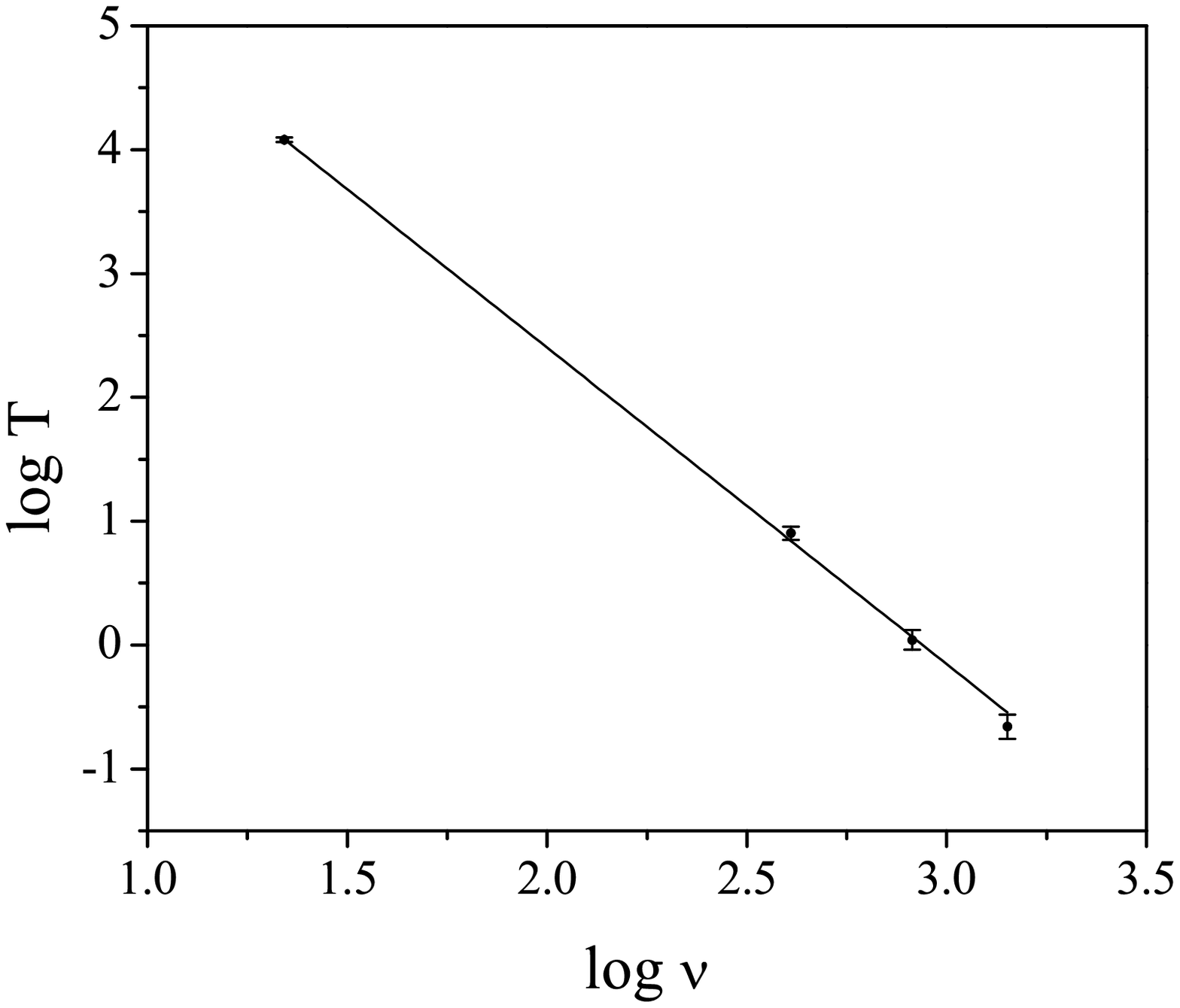}
\caption{The same as in Fig. 5, but for Loop II.}
\label{fig07}
\end{figure}

\begin{figure}[ht!]
\centering
\includegraphics[height=0.4\textheight]{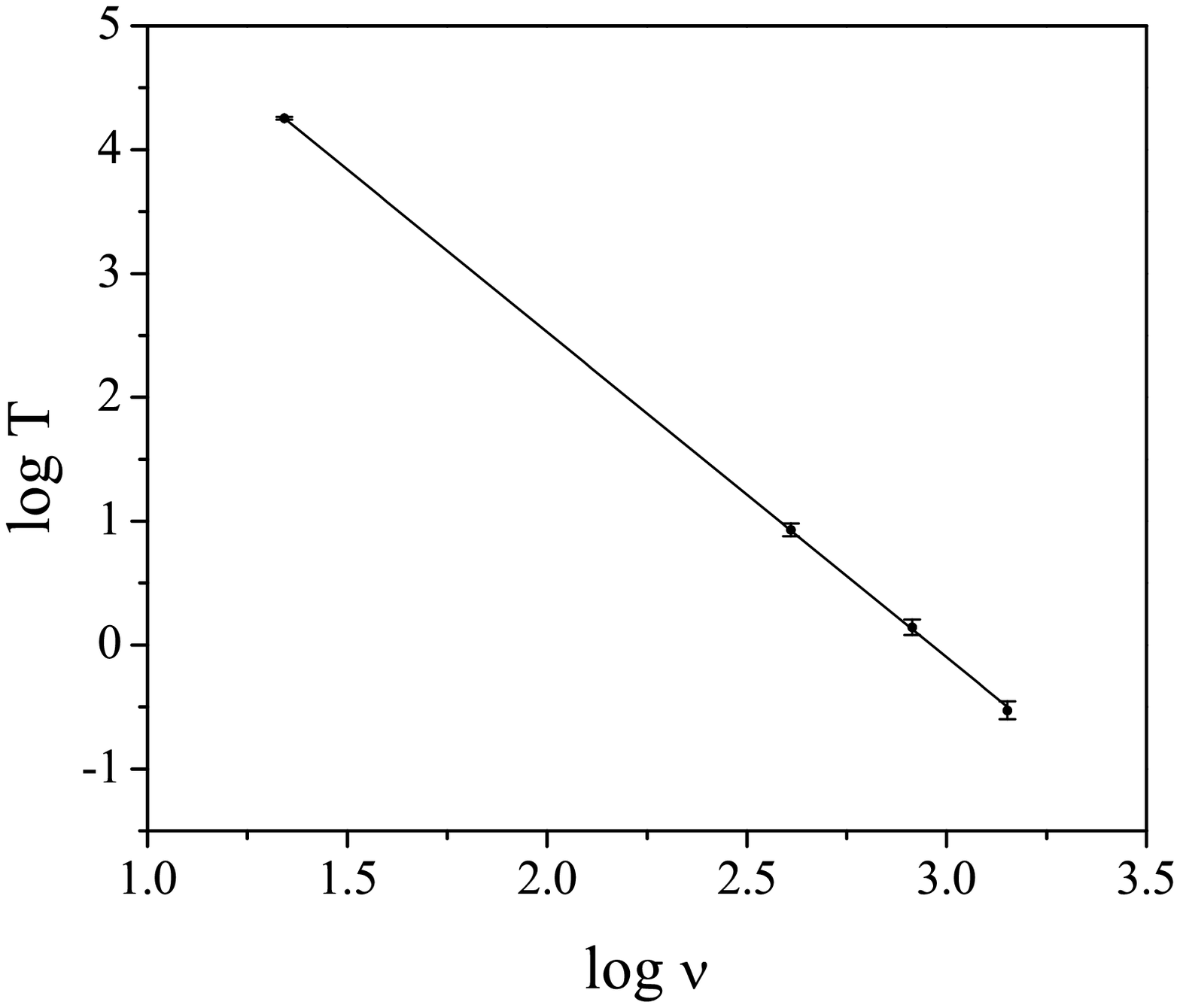}
\caption{The same as in Fig. 5, but for Loop III.}
\label{fig08}
\end{figure}

\begin{figure}[ht!]
\centering
\includegraphics[height=0.4\textheight]{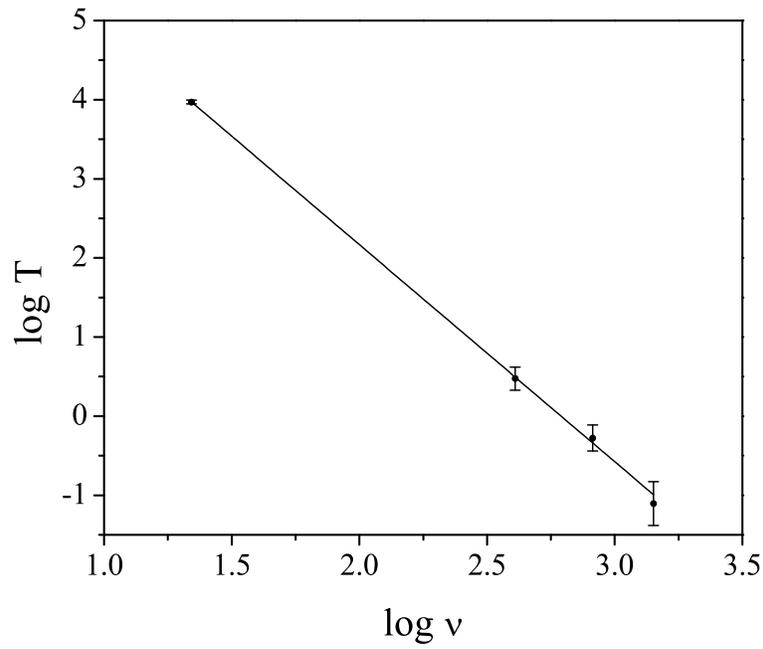}
\caption{The same as in Fig. 5, but for Loop IV.}
\label{fig09}
\end{figure}

\clearpage

\section{Conclusions}

i) In this paper, we have calculated the average brightness
temperatures and surface brightnesses of the radio-loops at 22 MHz.

ii) The values of spectral indices that we derived are slightly
shallower (but clearly non-thermal) than the values obtained in
Borka (2007). It should be expected effect because we added to the
spectra low frequency points at 22 MHz.

iii) We present the first radio-continuum spectra for the
radio-loops using average brightness temperatures at four different
frequencies.

\section{Acknowledgement}

We would like to thank the referee and handling editor for valuable
comments. This work is part of the projects 176005 ''Emission
nebulae: structure and evolution'' and 176003 ''Gravitation and the
large scale structure of the Universe'' supported by the Ministry of
Education and Science of Serbia.

\end{document}